\documentclass[12pt]{article}
\title{A practice-oriented overview of call center workforce planning}
\author{Ger Koole \& Siqiao Li\\CCmath bv \& Vrije Universiteit Amsterdam}

\usepackage{graphicx}
\usepackage{bbm}
   \newcommand{\E}{\mathbbm{E}}
\usepackage{a4wide}
\usepackage{xcolor}
   \newcommand{\red}[1]{{\color{red}#1}}

\begin{document}
\maketitle

\begin{abstract}
We give an overview of the practice and science of call center workforce planning, where we evaluate the commonly used methods by their quality and the theory by its applicability. As such this paper is useful for developers and consultants interested in the background and advanced methodology of workforce management, and for researchers interested in practically relevant science.
\end{abstract}
\section{Introduction}

Call centers are a fascinating area for stochastic modelling. In manufacturing most production is being done before the demand occurs, the product is lying on a shelf in a shop or a distribution center waiting for customer demand. In (non-urgent) health care production is smoothed in time to meet capacity: a patient makes an appointment with a health care provider at a moment that suits above all the provider. In aviation and hospitality demand is pushed by financial incentives towards low-demand time slots. Inbound call centers have in common with emergency health care that demand has to be met almost instantaneously by supply. And while a hospital has at least 15 minutes to prepare for the arrival of a trauma patient, a call center often has to answer a call within 20 seconds. And it can be life-saving, as is the case of an emergency call center. 

To be able to deliver this type of service planners have to deal with fluctuations, unforeseen (such as the variability of the Poisson process, or illness of employees, often called agents) and foreseen (such as intra-day and intra-week seasonality in demand). Call centers cannot react instantaneously to all fluctuations, and therefore have to schedule overcapacity. Designing the call center in such a way that little overcapacity is needed, and planning the right amount and types of overcapacity is the essence of workforce planning.  

Nowadays many call centers handle contacts through different communication {\em channels}, such as chat and email. However, inbound calls is often the most prominent channel. To give credit to the different channels the term {\em contact center} has been introduced. Few people however use it, thus a call center is most of the time a contact center mixing contact from different channels. A notable exception are call centers dedicated to outbound marketing campaigns. Through {\em predictive dialing} they deal with fluctuations in the fraction of calls that are answered and the speed at which this is done.
Quite a number of patents for algorithms can be found on Google Scholar.

Will there still be call centers in say a decade? We see a tendency for offering automated customer service by using for example AI in chat bots and call avoidance by for example improved web sites. Indeed, there is evidence that making calling unnecessary is the best customer service (Dixon et al.~\cite{DixonTomanDelisi}). And if people call, avoid that they have to make another call later on. 
Avoiding calls is also cheaper, and as most call centers are seen as cost centers, there is a strong incentive to reduce costs.
However, there is no evidence that the call center market is shrinking, on the contrary (Mazareanu~\cite{Mazareanu-marketsize}).
A possible explanation is the popularity of shared service centers which operate effectively as call centers (e.g., the human resources department at our university).
As such, we see a tendency across industries, from decentralised service to centralised service (operated as a call center, potentially {\em offshored} to a country with lower wages) to self-service.

\begin{figure}
\includegraphics[width=\linewidth]{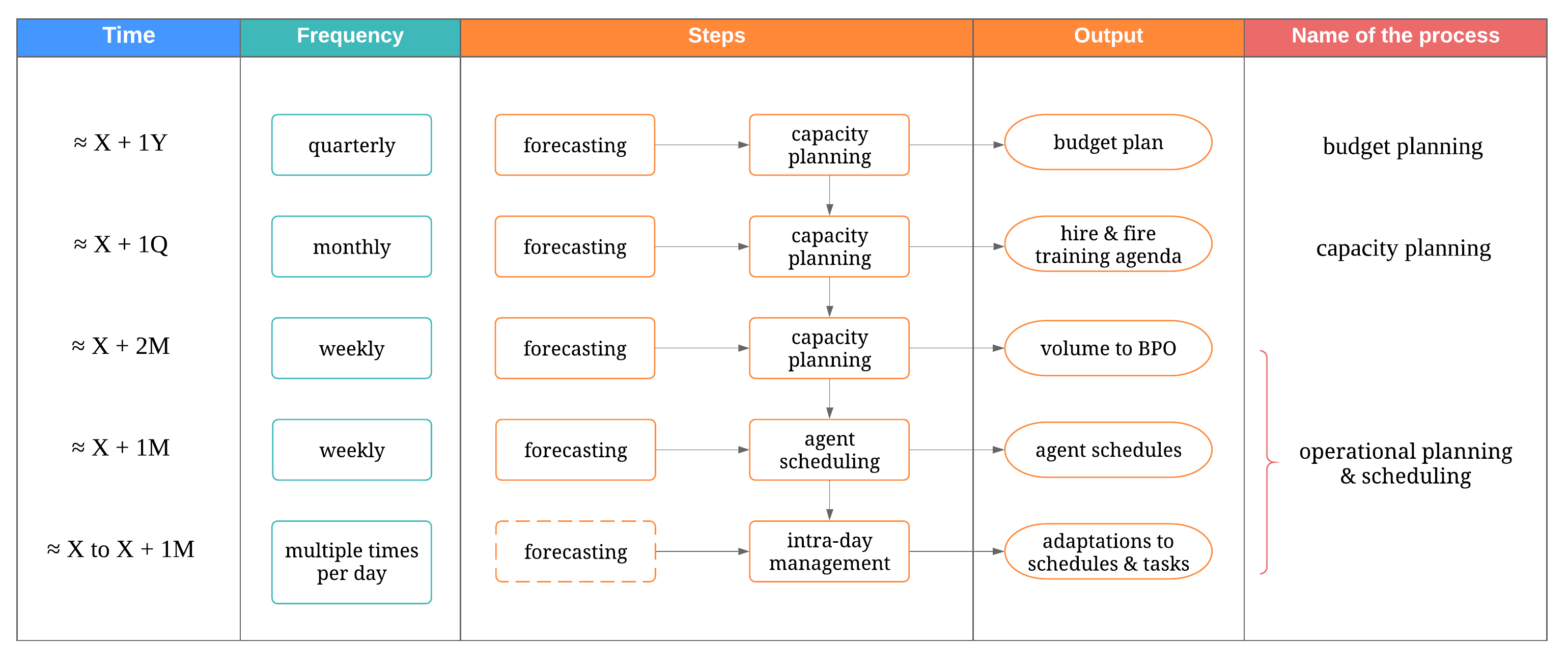}
\vskip-3mm
\caption{The WFM processes}
\label{proc-fig}
\end{figure}

This overview focuses on the practice of workforce management (WFM). (A better name would be workforce planning, but we will stick to the commonly used terminology.) As framework we use the different steps in the WFM processes. The three central planning processes are: budget planning, capacity planning, and agent scheduling. See Figure~\ref{proc-fig}. ``x" refers to the day of execution, 
``x+1Q" for example means 1 quarter before the day of execution. Note that many companies use {\em business process outsourcers} (BPOs) to handle (parts of) their call volume. To allow them to prepare for their job forecasts or required staffing levels are communicated at multiple moments in time. 

As can be seen from the figure every step starts with forecasting. An exception is intra-day management: the forecast is rarely updated after the agent schedule is made, although that would likely result in an increased accuracy. There are three processes:\\ 
- the long-term budget process, which is input for the corporate management which sets the financial boundaries;\\
- the tactical capacity planning process, where decisions concerning the agent pool are made, mostly the hiring of new agents and the training of new skills;\\
- the short-term operational planning process, which starts with deciding which volume goes to the external partners, and then consists of agent scheduling (which need to be communicated a few weeks in advance) and finally intra-day management which is done at the day itself. 

Depending on the particular call center the situation might be slightly different, and smaller call centers with stable volumes might not execute the long-term steps explicitly. Note also that this scheme is biased towards the European situation with its strict labor laws forcing call centers to schedule carefully and publish schedules well in advance.

Next to the three processes from Figure \ref{proc-fig} call centers have two more less explicit processes: a long-term, ad-hoc process which is about improving the overall design of the call center: the shift structure, the way forecasts are made, opening times, channels that are offered, etc. This process also designs all underlying processes. Finally, there is the real-time routing, the assignment of customer contacts to agents, which has a big impact on the performance. This is automated and part of the telephony/omnichannel switch. Although this could also benefit from updated forecasts and other real-time information this is rarely used.

Note also that we left out the connections with other departments. Forecasting for example takes input from marketing and sales to obtain the dates of marketing campaigns and sales forecasts, and the budget plan is used in negotiations with higher management to set the final budget. Furthermore, the processes are not always as linear as they seem: there might be interaction between forecasting, scheduling and marketing about the feasibility of marketing campaigns; capacity planning might lead to adaptations to the budget, etc.

WFM has a supporting role in a call center. It helps achieve the goals of the three stakeholders, customers, employees and management: give good service by satisfied employees at a reasonable price. Good service is usually defined by service level agreements (SLAs) which serve as constraint in all WFM steps. Employee satisfaction is represented for example in the types of possible shifts and fairness among agents that routing rules achieve. The financial side is the main reason for having the budgeting cycle, and budget is discussed regularly to see if there are any exceptions.

In the next sections we discuss one by one the different steps of the WFM processes. We conclude with a section on overall call center design principles. Papers and other relevant sources of information are cited where appropriate. We do not try to be complete with respect to the literature, we focus on what we consider to be most relevant to practice. Papers inspired by call centers but of little use to its operations are left out.

We end this introduction by citing some general call center WFM references. The following are academic overviews: Gans et al.~\cite{GansKM}, Avramidis \& l'Ecuyer~\cite{AvramidisE05}, and Ak{\c s}in et al.~\cite{AksinAM}. More practitioner-oriented text books are Cleveland \& Mayben~\cite{ClevelandM} and Koole~\cite{Koole-CCWFM}.

\section{Forecasting}%%%%%%%%%%%%%%%%%%%%%%%%%%%%%%%%%%%%%%%%

Call center forecasting concerns the prediction of call volumes at the interval level, usually per quarter. As can be seen from Figure \ref{proc-fig}, forecasts are needed at all time-levels, from a few intervals in advance to long-term forecasts years ahead. These forecasts should take all factors that influence volume into account: long-term trend, intra-year, intra-week, intra-day seasonality, and events such as holidays, marketing actions, and IT problems of products and call center systems. An important task of the forecaster is to explain what he or she predicts, thus it is important that the forecasting method is transparent such that the forecaster can say something like: ``Next week on Monday we have 2000 calls more than last week. There is a marketing campaign with an expected impact of 3500, but the base level is 1500 lower because of the holidays." Managers won't allow decisions to be made on forecasts purely based on black-box forecasting, they want the reasons behind a prediction. Note that, in contrast with the linear processes described in Figure \ref{proc-fig}, there might there be interaction between forecasting and planning: events such as marketing campaigns might be planned on the basis of agent availability.

In practice most call centers either use a self-made spreadsheet or judgemental forecasting. Forecasting is done first at the daily level, for example by a simple decomposition approach that adds the increase over a year to last year's volume, in a formula with $h$ historical volumes, $\hat y$ the forecast, and $w$ and $y$ time periods of a week and a year:
\[\hat h_t=h_{t-y}{h_{t-w}\over h_{t-y-w}}.\]
This forecast is adapted using estimations the impact of events on $t$, $t-w$, $t-y$, and $t-y-w$. It can be made more sophisticated by predicting weekly volumes and intra-week profiles and by estimating the yearly increase by averaging over multiple weeks. Some scheduling tools offer forecasting functionality but rarely more advanced. Very few call centers employ advanced forecasting methods.

In the literature many methods have been proposed and applied to call center data. 
Taylor~\cite{Taylor-ccfc}, Jalal et al.~\cite{JalalHK-ccfc}, Antipov \& Meade \cite{AntipovM-ccfc}, Weinberg et al.~\cite{weinberg2007bayesian}, Ibrahim et al.~\cite{ibrahim2013forecasting} and Huang et al.~\cite{HuangJDZ-ccfc} are some of them.
An elaborate overview is given in Ibrahim et al.~\cite{IbrahimYES-forecasting-survey}.
All these models, using different (combinations of) algorithms from statistics and AI, are successful in the context in which they are described, but there is no consensus on which method is preferable in the most common situations. 
Although various lead times (i.e., times in advance for forecasting) are considered when the proposed forecasting approaches are compared, most works focus on short term forecast such as daily forecast or intra-day forecast without considering intra-year seasonality (impact), which cannot be ignorant in practice. 
Also, most methods lack features that are required to handle all drivers of call center volume.
Especially event handling is often lacking. Some exceptions are Aldor-Noiman et al.~\cite{aldor2009workload}, Antipov \& Meade~\cite{AntipovM-ccfc} and Soyer \& Tarimcilar~\cite{soyer2008modeling}, which incorporate the effect of events (e.g., marketing strategy and special calendar effects) as exogenous variables in their mixed-Poisson arrival count models.
A good method to deal with all aspects is described in Hyndman~\cite{Hyndman-events}, using smoothing methods as described in Hyndman \& Athanasopoulos~\cite{HyndmanA} and a separate regression for events using dummy variables. 
Also a regression model with a polynomial for the trend and events modeled in the same way works quite well (Koole~\cite{Koole-CCWFM}). 
Because call center actuals are multiplicative in their components (Ding \& Koole~\cite{DingKoole-WAPE}) it is advisable to use Poisson regression, i.e., use the regression on the logs of the actuals.

Decomposition methods, by which we mean methods that determine the factors that influence volume one by one, only work in a multi-pass setting because of the dependences of the underlying variables. For example, the occurrence of outliers can only be determined if you know the seasonalities. But the seasonalities can be better estimated if the events and outliers are filtered out. While some form of decomposition is commonly used by forecasters in practice, multi-pass methods are rarely used and neither studied in the literature.

Forecasting errors have to be measured using some criterion. It is an important task of the forecaster to report and explain forecasting errors to management. Therefore the error measure should be easy to interpret. The WAPE (weighted absolute percentage error) is a good candidate, also because it is less prone to outliers in small volumes than the MAPE, and because the WAPE is linear in the intra-day management costs (\cite{DingKoole-WAPE}). 

Call center arrivals can well be modeled as coming from an inhomogeneous Poisson process (\cite{kim2014call}), thus we forecast the arrival rate. This gives a minimal error, which is, in terms of the APE (absolute percentage error) equal to $\E|N_\lambda-\lambda|/\lambda$, with $\lambda$ the forecast and $N_\lambda\approx\textrm{Poisson}(\lambda)$.
A formula is given in Crow~\cite{Crow}, the (very good) normal approximation is $\sqrt{2/(\lambda\pi)}$. Taking a weighted average of the minimal APE for each interval gives the minimal WAPE for any time period consisting of multiple intervals that is to be forecasted.

It has been observed that call center data is overdispersed with respect to the Poisson distribution (see Jongbloed \& Koole~\cite{JongbloedK} and Avramidis et al.~\cite{avramidis2004modeling}). However, this simply means that there is a considerable error, which might well be largely explained by adding more features such as events. Evidently, the risk of overfitting is present, but a good forecasting model taking all relevant features into account can reduce the overdispersion enormously. These features include, next to events such as marketing actions, special days, and IT issues, also the weather and other time series such as sales forecasts. The weather, especially the derivative of the temperature, has an impact on call volume: the first day of nice weather sees a decrease in calls in countries such as the Netherlands. To forecast call volume using the weather you need a good weather forecast. Therefore including the weather only works for short-term forecasting. A simple implementation would be to include the first day with nice weather as a recurring event of which the impact can be determined from previous days with nice weather. Depending on the granularity, horizon and characteristics of the call center a considerable WAPE on top of the minimal WAPE might still remain. Although 5\% is considered to be the golden standard, this varies wildly in practice and is regularly much higher than 5\%. 

In the urge to explain the forecast forecasters, and especially their managers, like to include time series such as sales in their forecast. Forecasts made this way are called ``ratio forecasts", because a (known) fraction of new customers calls. However, again a forecast of the external variable is needed, while the trend most of the times show considerable collinearity with the sales. Furthermore, the fraction might change.  Therefore it is questionable whether including a sales forecast improves the forecast. Testing it is the only way to find out, and often it is indeed not the case. Adding a variable such as a sales forecast is useful if it contains information not yet contained in the call volumes, such as qualitative opinions. This is often the case with long-term forecasts, made for budgeting reasons or capacity planning. An additional advantage is that it explains the forecasts and also its errors.

Note that these errors will be considerable, especially for long-term forecasting (Makridakis~\cite{Makridakis}). 
Therefore, the question should not be whether the forecast is reliable, but if we have enough flexibility to deal with the inevitable error. Call center managers recognise the importance of flexibility, but hardly ever make the connection between forecasting errors and the amount of flexibility required. This should be part of {\em capacity planning}, the long-term determination of the required capacity. 

\begin{figure}
\centering
\includegraphics[width=0.75\linewidth]{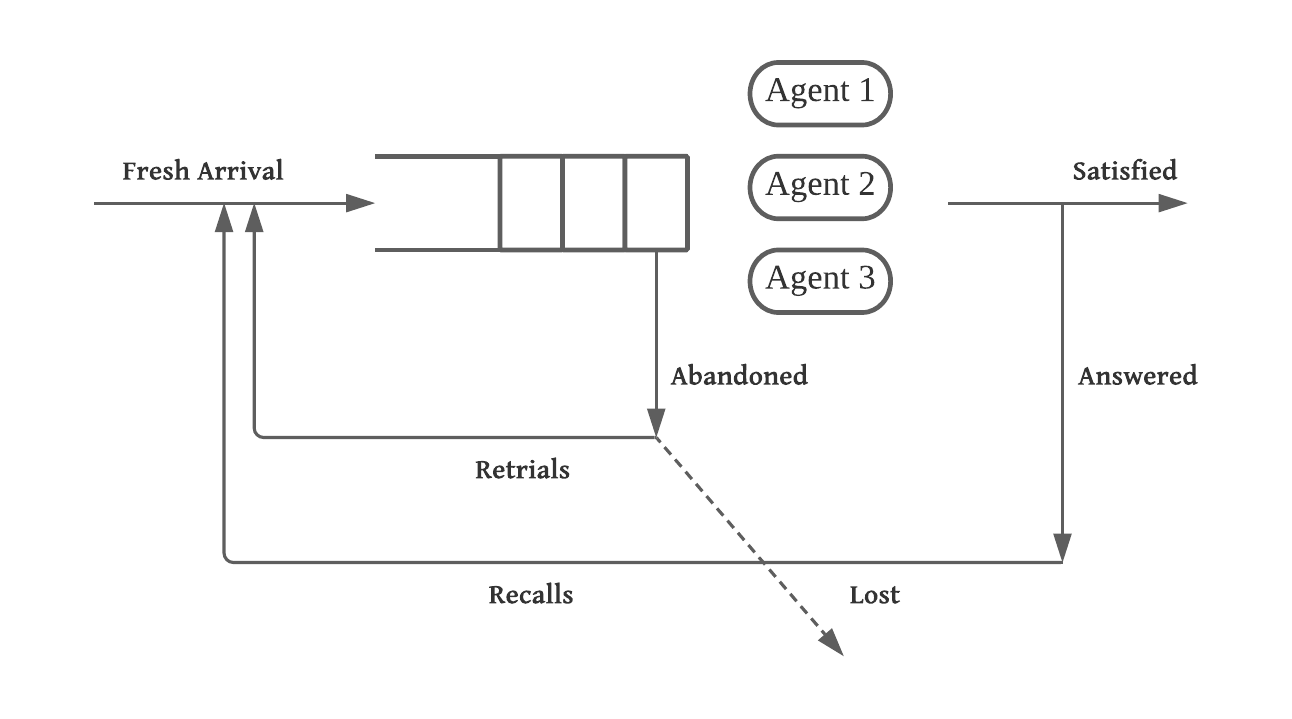}
\vskip-3mm
\caption{Retrials and recalls}
\label{rr-fig}
\end{figure}

Forecasting is often done for the total number of offered calls. However, this includes retrials: callers who abandoned earlier and called again later, see Figure \ref{rr-fig}. Data analysis shows that retrials often occur shortly after the first attempt, usually within the same day (Ding et al.~\cite{DingKooleMei-truedemand}). Usually one knows the numbers of connected and abandoned calls, but the fraction of retrials is not known, unless the callers can be identified. An empirical study of retrial behavior can be found in Hathaway et al.~\cite{hathaway2017queue}. Ding et al.~\cite{DingKooleMei-truedemand} propose a statistical method to determine the ``fresh" volume by using the retrial percentage as a variable in the forecasting model. In practice taking the average between the offered and handled numbers of call often works well, corresponding to 50\% retrials. Note that there are also {\em recalls}, callers who call a second time to get advice. Recalls add to the call volume and therefore to the workload, but they are also a very important driver of customer dissatisfaction (Dixon et al.~\cite{DixonTomanDelisi}). Reducing it however is outside the scope of WFM.

After determining daily volumes they have to be drilled down to the intra-day volume. Typically forecasters will base themselves on 
what they consider to be similar days as the one they are about to forecast: same day of the week, not too long ago, similar events. Then they take the average of the {\em profiles}, the normalised volumes, and multiply that with the daily forecasts. However, this method leads to considerable overfitting, the average profile often shows quite some variability.
Much better results are obtained by using the fact that you expect neighbouring intervals not to vary that much, by fitting a polynomial or a smoothing spline. This proves to work quite well (Bakker et al.~\cite{BakkerCKLR-splines}, Soyer \& Tarimcilar~\cite{soyer2008modeling}, Channouf \& L'Ecuyer~\cite{channouf2012normal}). 

In this section we fully focused on forecasting inbound calls, but the same methods apply to other forms of customer contact such as email and chat. Next to that, other parameters needed for scheduling and capacity planning require forecasting as well. Examples are handling times and form of {\em shrinkage} such as sick leave. There are some differences, and usually not the same granularity is required (sick leave is a parameter at week level), but overall the same methods can be used. Note that handling times also show fluctuations during a day. Aktekin~\cite{aktekin2014call} and Tan \& Netessine~\cite{tan2014does} observe that the time of the day affects agents' handling times. 
For example, they may speed up their service when the call center is busy.

\section{Staffing}%%%%%%%%%%%%%%%%%%%%%%%%%%%%%%%%%%%%%%%%

Agent scheduling concerns the construction of schedules such that, amongst other objectives, SLAs are expected to be met to the extent possible. Commonly used SLAs are 1 minus the tail of the waiting time distribution (also referred to as the service level (SL), often taken as 80\% answered within 20 seconds) and the expected waiting time (the {\em average speed of answer}, ASA). With the SLA as constraint the minimum required staffing in every interval is determined, sometimes explicitly, or implicitly in the scheduling algorithm. If it is done explicitly, and then given to the scheduling algorithm, then it is often done by the forecasters and even called {\em workforce forecasting}. We will call it {\em (safety) staffing}, as it entails planning overcapacity to deal with fluctuations in workload. Staffing is probably the best-studied part of WFM, and the starting point of many scientists interested in WFM, explaining why many queueing scientists (used to) work on call centers. 

We will make a difference between single and multi-channel and single and multi-skill operations. Staffing is done at the interval-level, usually 15 minutes. Even though agents can often handle multiple skills and/or channels, they are often scheduled during one or more interval to work on a single skill and/or channel. We will first look at staffing for these single-skill single-channel operations, starting with inbound. Then we will look at staffing in a {\em blended} multi-channel environment and in the presence of {\em skill(s)-based routing}, where in real-time a contact from the optimal channel or skill is being pushed to the agent.

It is commonly assumed that arrival rates and numbers of agents are stepwise constant functions, constant during each quarter. This is motivated by the fact that arrival rates are expected to change little during each quarter, and that schedule changes are only possible at the quarter. In this situation the so-called SIPP approach (Green \& Kolesar \cite{GreenKpsa}) is an obvious choice: you assume stationary in each interval, and use a stationary queueing model. The $M|M|s$ or Erlang C model is most commonly used in practice. Allowing for customers to abandon is a relevant feature that improves the approximation. The model including abandonments is commonly written as $M|M|s+G$, with $+G$ denoting the generally distributed patience. In the case of exponential patience we call it Erlang~A. Seminal work on these models was done by Palm \cite{Palm} and Baccelli \& H\'ebuterne~\cite{BaccelliH}, Zeltyn \& Mandelbaum \cite{ZeltynMandelbaum} gives simple formulas for the $M|M|s+G$ using integrals over the patience distribution. Sze~\cite{Sze} added retrials. 

Compared to the Erlang C the Erlang A has one more parameter: the patience distribution or its expectation, depending on the exact model used. The waiting time of a customer is the minimum of its patience and the time to service, thus the patience is a censored variable. The famous Kaplan-Meier method (\cite{KaplanMeier}) can be applied, leading to results as in Brown et al.~\cite{BrownGMSSZZ} who did a thorough analysis of call center data. Note that in practice patience is usually underestimated because practitioners often only look at the abandoned calls. However, taking an expected patience of 5 or 10 minutes is already much better than applying Erlang C. You can also use the patience as a tuning parameter, but then you need data on the achieved service levels. 

A handful of papers focuses on the patience distribution and how it affects call center performance and staffing decisions. For example, Mandelbaum \& Zeltyn~\cite{mandelbaum2004impact} study the impact of various patience distributions on $M|M|s+G$ queues. They observe approximate linearity between the probability to abandon and the average waiting time when there is a low to moderate abandonment rate. Roubos \& Jouini~\cite{roubos2013call} empirically show that the hyper-exponential distribution is an accurate representation of the patience distribution. Aktekin \& Soyer~\cite{aktekin2014bayesian} conduct Bayesian analysis built upon different families of distributions. Ye et al.~\cite{ye2020hazard} estimate the hazard function of customer patience time with a nonparametric approach.
It is worth mentioning that Whitt~\cite{Whitt-ms05-engineering} show that the patience distribution has a bigger impact than the service time distribution for the same expectations. 

Erlang A, compared to Erlang C, gives also the possibility to include the abandonment rate in the SLA, both separately, the fraction of abandonments needs to stay below for example 5\%, or implicitly in the SL (\cite{JouiniKooleRoubos}). It is common in science to use the {\em virtual waiting time}: the time an arbitrary customer would have to wait if her patience were $\infty$. However, this measure is not measured in a call center, thus the performance cannot be verified. In practice other definitions are used, for example the fraction of all calls being answered within the {\em time to answer}. For definitions and ways to compute the SL for different definitions see  Jouini et al.~\cite{JouiniKooleRoubos}.

It is interesting to note that delay announcements, which provide estimates of the waiting time, influence the patience.
Psychologically, ``uncertain waits feel longer than known finite waits" \cite{maister1984psychology}. 
On the other hand, the delay announcement can also induce some customers to balk or abandon earlier, leading to peaks in abandonments. This, in its turn, influences the waiting times \red{\cite{feigin2006analysis, armony2009impact, jouini2011call, ibrahim2009real, ibrahim2011wait}}. Ak{\c s}in et al.~\cite{akcsin2017impact}, as one of the latest work on this topic, applied a series of Cox regression to a bank call center data with delay announcement messages every 60 seconds and revealed that both the content (i.e., how detailed waiting time information is offered) and the sequence (i.e., positively/negatively change of the waiting situation) announcement messages, the congestion levels of the call center and the characteristics of customers have a statistically significant impact on abandonment behaviour, however, due to the complexity, no staffing decisions considering delay announcements have been studied yet.

SIPP combined with an Erlang model is the most commonly used method in practice. However, there are a number of problems with such an approach. We will discuss them one by one.

In the first place, the queue is not in a stationary situation at the beginning of each interval. Depending on the parameters of the previous intervals, you might for example expect a backlog. There are a number of methods available with this, of which, according to Babat~\cite{Babat-MtMst}, the {\em stationary backlog carryover} (SBC) approach by Stolletz performs best (\cite{Stolletz-SBC}). 

In the second place, the SIPP predicts expected performance. If you schedule using SIPP at the level of your SLA, then in roughly 50\% of the cases you won't reach your daily SLA. The error can be quite big (Roubos et al.~\cite{RoubosKooleStolletz}). In practice, this is unacceptable. A solution might be to look at quantiles of the distribution of the service level (\cite{RoubosKooleStolletz}), but usually the problem is solved by intra-day management. 

In the third place, on top of the transient effect we just discussed, there is uncertainty about the arrival rate, the overdispersion we found in the previous section. Scheduling according to the expected rate is suboptimal, Ding \& Koole~\cite{DingKoole-WAPE} propose a method that integrates the intra-day adaptations into the staffing step, leading to a newsvendor-type staffing method. Whitt~\cite{whitt2006staffing}, Steckley et al.~\cite{steckley2004service}, and Liao et al.~\cite{liao2012staffing} also incorporate uncertain arrival rates into staffing planning.

In the fourth place, there are many factors that influence performance that are not modeled by Erlang C nor Erlang A, such as the impact of short unscheduled breaks and the behavior of agents under longer periods of high workload. Only recently the first attempt to validate the Erlang models based on realized service levels was undertaken (Ding et al.~\cite{DingLKYMS-validation}).
By studying agent data together with call data it was indeed found that breaks have a huge impact on performance. One way to solve this is to move to a statistical/machine learning approach that takes all features into account, as in Li et al.~\cite{LiKooleYuceCatanese-MLstaffing}. This is even better than simulation because implicitly the behavior of the agents is taken into account; to use simulation it has to be modeled explicitly which is hard because of differences between agents and the lack of knowledge on how and when for example breaks are taken.

The widely used Erlang97 Excel add-in (Bromley~\cite{Bromley-Erlang97}) also has the option to compute abandonments. It is based however on a waiting-time quantile of the Erlang C, thereby making two errors: it does not model the fact that Erlang A generally has a better SL than Erlang C because some customers leave the queue, and it assumes the patience is the same for all customers (van Eeden et al.~\cite{EedenHilstKoole-Erlang97}).

It is hard to obtain qualitative insights from the Erlang formulas, for example how they behave when you increase scale. {\em Square-root staffing} does. For $\lambda$ the arrival rate and $\beta$ the average handling time, it says that staffing should be at $\lambda\beta+\alpha\sqrt{\lambda\beta}$ with $\alpha$ a parameter depending on the SL only. The square root can intuitively be interpreted: if you add 2 i.i.d.\ r.v.s then the standard deviation is multiplied by $\sqrt2$. The same holds for the safety staffing, because it is there to handle fluctuations in load. This clearly shows the economies of scale which is one of the reasons why we want agents to be multi-skilled. It also shows decreasing returns, as $\lambda\beta+\alpha\sqrt{\lambda\beta}$ is concave in $\lambda\beta$, which tells us that not all agents need to be multi-skilled. 
Halfin \& Whitt~\cite{HalfinW} introduced this {\em Halfin-Whitt regime} in which the load increases but the delay probability is held constant. Since then many papers have studied this regime in many different variants, see \cite{reed2009g, harrison2004dynamic, braverman2020steady, gamarnik2012multiclass} for recent ones. Unfortunately these ideas are very little used in practice.

Next to inbound a variety of other channels are used. They can be divided into synchronous and asynchronous communication. Email, webforms and old-fashioned mail and fax are asynchronous. Usually the time-to-answer is multiple hours or days, at least multiple intervals. This means that fluctuations have to be dealt with by flexibility in scheduling, not by safety staffing. A noteworthy synchronous channel is {\em chat}. The difference with inbound from the point of view of WFM is that a chat agent can do multiple chats in parallel, usually 2 or 3. When all agents are saturated customer wait in the queue, just like an Erlang system. The parallelism increases the efficiency, when an agents answer one customers the other(s) can formulate their responses. However, it makes the total time per chat longer: sometimes a customer has to wait for a chat to be available. Quantifying the durations are challenging, but can then be used to extend the Erlang models for chat. See Koole~\cite{Koole-CCWFM} for more details on an implementation.

Moving decisions to a later moment when better information is available is a general principle. 
One way to do this is to move the decision which type of task to do from the schedule to the routing.
In a multi-channel environment this leads to {\em blending}, in a multi-skill environment to {\em skill-based routing} (SBR).
We discuss how staffing can be done in these environments.

Blending is usually executed by blending synchronous and asynchronous channels, such as inbound and email or outbound. When the asynchronous channel can be interrupted to deal with priority with inbound, then staffing is easy: inbound is staffed as discussed, and the overcapacity with respect to the expected load is filled with email. Things get more complicated when the asynchronous channel cannot be interrupted as in the case of outbound. This case has been studied in Bhulai \& Koole~\cite{BhulaiK03ieeeac} and Gans \& Zhou~\cite{GansZ03}. Both the routing policy and, implicitly, staffing has been determined. No papers discuss blending of asynchronous channels such as inbound and chat. A simple policy could be to assign an agent to the channel with the longest waiting customer. To use as little agents as possible for chat, chats should be assigned to agents already handling chats but who are not yet saturated. 

Now we move to SBR.
Because of the lack of closed-form formulas for the stationary situation, simulation is the only viable option for SBR, apart from some approximations based on models without waiting (Chevalier \& Tabordon~\cite{chevalier2004routing}, Pot et al.~\cite{PotBK}) and based on fluid models considering abandonment targets as the Quality of Service constraints (Gurvich et al.~\cite{gurvich2010staffing}, Bodur and Luedtke~\cite{bodur2017mixed}, Bassamboo et al.~\cite{bassamboo2006design}).
But why run a long-term simulation to find stationary behavior when it is easier to do a short-term transient analysis, for example of a day?
For this reason most studies tackle right away the scheduling problem, resulting in little literature on multi-skill staffing alone. 
Some notable exceptions are Cezik \& l'Ecuyer~\cite{CezikE}, Gurvich et al.~\cite{gurvich2010staffing}, Harrison et al.~\cite{harrison2005method} and Chan et al.~\cite{chan2014chance, chan2016two}.

\section{Agent scheduling}%%%%%%%%%%%%%%%%%%%%%%%%%%%%%%%%%%%%%%%%

Agent scheduling is the operational process in which agents get assigned to shifts and activities during these shifts. 
Activities include the channel and/or skills they have to work on, but also paid breaks, meetings, trainings, etc.
Next to the routing, which is part of the telephony/omnichannel switch, it is the part of WFM that is most often supported by specialized software.
There is a wide choice of software vendors, the bigger ones include Genesys, Injixo, Nice, Teleopti, and Verint. 
See for example \cite{TrustRadius-WFMsoftware} for a list.
However, little is known about their exact workings, Erlang C and simulation are used, the latter often leading to very long run times.
Fukunaga et al.~\cite{FukunagaHFAMN} give some details about Verint (called Blue Pumpkin at the time). 
Smaller call centers, and also the ones with less scheduling issues (for example, because they are only open during business hours), often schedule using a spreadsheet.
Agent scheduling in its generally is hardly studied in the literature: usually unpersonalized shifts are determined, without activities within the shifts, which is actually shift scheduling.

\begin{figure}
\centering
\includegraphics[width=0.75\linewidth]{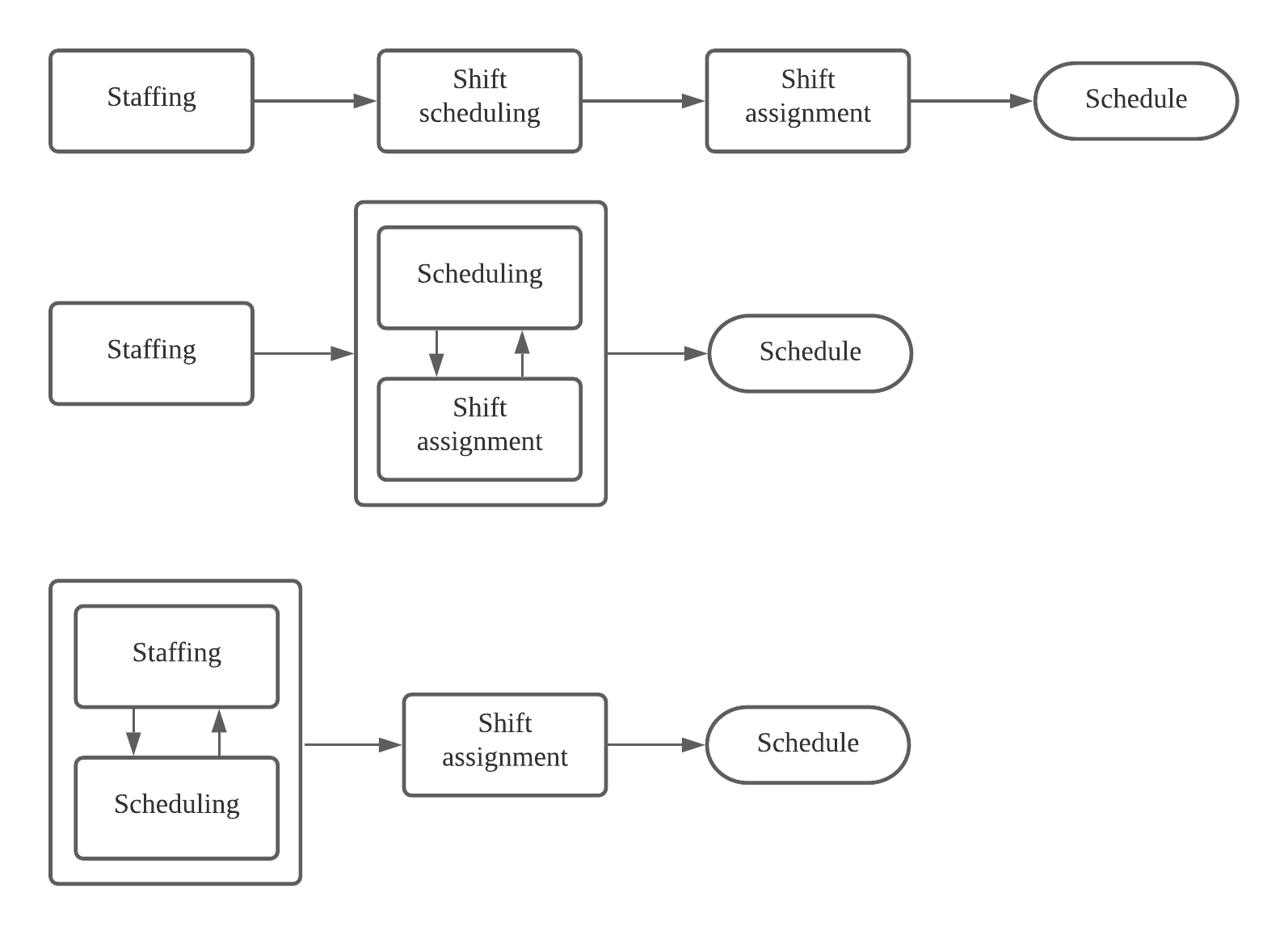}
\vskip-3mm
\caption{Short-term operational planning process}
\label{ss-fig}
\end{figure}

In its simplest form agent scheduling consists of three steps, as illustrated in the top part of Figure \ref{ss-fig}: for each interval the required staffing levels is determined (e.g., using an Erlang formula), the most efficient way to cover the staffing needs by the available shifts is determined (potentially using integer linear programming (ILP), and these shifts are assigned to agents in some way (for example, by letting them choose in the order of seniority). The first to formulate a solution for the covering problem of the second step was Dantzig in \cite{Dantzig-toll-booths}, in which he considered toll booths at a US bridge.

There are various reasons why such an approach is highly suboptimal and even infeasible. 
Often employees have different types of contracts, therefore in step 2 different groups of shifts should be identified, otherwise no match between agents and shifts can be made. Furthermore, many agents have personal preferences. Satisfying as much as possible is crucial for employee satisfaction, making that the schedule should be made at the individual level, integrating step 2 and 3, as in the middle part of Figure \ref{ss-fig}. Dealing with these personal preferences is evidently part of WFM software, but hardly studied scientifically.

A much better studied subject is the integration of step 1 and 2, as in the lower part of Figure \ref{ss-fig}. The reason for combining them is that the staffing levels of step 1 are hard to cover with shifts, leading to considerable overstaffing.
Often SLAs are formulated at the daily level, thus SLs are allowed to fluctuate a bit, certainly if that leads to more efficient schedules and if the daily constraints are met.
Integrating step 1 and 2 makes the optimization problem highly non-linear. 
It can be rewritten as ILP but a the cost of having many binary variables. 
Add to this the fact that we should schedule at the weekly level (necessary because of constraints on the schedules related to numbers of working days per week and start times), then we are stuck with heuristics such as local search. 
In multi-skill settings we need simulation to get reliable evaluation of possible solutions, leading us to simulation-optimization with stochasticity on the SL constraints, problems which are known to be notoriously difficult. 

In a single-skill situation, Koole \& van der Sluis~\cite{KooleSluis} uses SIPP and shows that under a very simple shift structure, a suitable local algorithm can find optimal schedules. In a transient single-skill setting, Atlason et al.~\cite{AtlasonEH} use simulation to generate cutting planes used in the shift optimization module. 
 Liao et al.~\cite{liao2013distributionally} also use simulation. They combine stochastic programming and robust optimization to work out scheduling with uncertain arrival rates. Robbins \& Harrison~\cite{robbins2010stochastic} solve a stochastic scheduling problem to minimize the combined cost of agents and missing QoS targets. Later, Gans et al.~\cite{gans2015parametric} considered a two-stage scheduling problem which allows adding and removing agents based on updated forecasts at midday. 
 
In a multi-skill situation, Pot et al.~\cite{PotBK} \& Bhulai et al.~\cite{BhulaiPK} use an overflow approximation for SBR, similar to Chevalier \& Tanbordon~\cite{ChevalierT}, to build a multi-skill scheduling algorithm. Bodur and Luedtke~\cite{bodur2017mixed} also replace the abandonment level by the approximating formula and solve a two-stage stochastic programming for scheduling with Benders decomposition. A main drawback of these approximations is unrealistic assumptions of idealistic fluid routing policy. Moreover, service levels cannot be approximated.

Again leveraging on simulation, Cezik \& l'Ecuyer~\cite{CezikE} extends the approach of Atlason et al.~\cite{AtlasonEH} to multi-skill staffing. Avramidis et al.~\cite{AvramidisCGEP} extends the cutting plane method to solve scheduling problem over a day (i.e., multiple periods). Running times however are very long.

The current state-of-the-art is Li et al.~\cite{LiWangKoole-scheduling} which uses machine learning (ML) to speed up the simulations. This makes it possible to solve industrial-size weekly multi-skill multi-channel problems in several minutes. 
Solving the same problem without using ML takes much longer, see Li et al.~\cite{LiKooleJouini-omnisim}.
Note that it is inevitable that the SL fluctuates, because of our daily SL objective. In call centers planners spend long hours adapting schedules manually to get smooth service levels, evidently a useless and expensive practice.

Note that all these problems consider {\em shift} scheduling: they determine shifts, but do not determine the activities within the shifts. This adds a layer of complexity far beyond the current state-of-the-art, but it is required in the operations and done by WFM software. On the other hand, it can be argued that the activity assignment should be done at the routing level, although some activities (such as meetings) need to be planned in advance. The fact that these methods are not at the level of agents but at best at agent group level makes them better suitable for capacity planning, which is the subject of the next section.

\section{Capacity planning}%%%%%%%%%%%%%%%%%%%%%%%%%%%%%%%%%%%%%%%%%%%%%%%%%

Capacity planning is the holy grail of WFM. To be able to do long-term planning you have to take into account how you all the shorter-term processes, thus all decisions at all levels impact capacity planning. On the other hand, it does not have to be done at the same level: while agent schedules need to be determined at the quarter level, capacity planning can often be done at the week level.

Let us first consider capacity planning used for budgeting. The long-term forecast is an essential element for this process. Based on the forecast schedules could be made, just as for agent scheduling. Then the costs of these schedules could be determined leading to the budget. However, apart from some practicalities such as the lack of information on agents still to be hired, runs are often too long to compute the multi-year horizon needed for the budget, especially because Excel has to be used as the planning tool is not appropriate for this kind of calculation.
A simple fast calculation is to estimate the budget proportional to the volumes: if the volume increase by $x$\% then the costs will also increase by $x$\%. Of course we make an error: costs are not linear in the forecast, but for small changes the error is expected to be small, probably much smaller than the forecasting error. More advanced methods, such as an ML model to estimate costs based on the forecast and other parameters, have been successfully used in practice.

More complicated are decisions related to the hiring (and perhaps firing) of agents and decisions about the training of new skills to existing agent. Hiring and training new agents is a lengthy process that can easily take 3 months or more, thus the capacity has to be planned well in advance. In the simplest case it is just deciding how many agents are needed, but often there are choices in types of contract and initial skill sets. To determine which types of agents to hire shift scheduling has to be done, over a longer period, starting from the current pool of agents, taking agents resignations and {\em shrinkage} into account. Shrinkage are all activities that avoid agents from being available for phone work (or other types of contacts), from holidays and illness to meetings and paid breaks. The operational schedule should take activities like meetings and short breaks into account, capacity planning all of them. Note that they are sometimes unpredictable, such as illness, and sometimes planable, such as when agents go on holidays or when meetings take place. Both types complicate capacity planning.
Many call centers do capacity planning in a grossly simplified way by replacing all randomness and advanced calculations by fractions as explained in the previous paragraph on budgeting.
Probably even more call centers use no calculations at all but make rough estimations, potentially making big errors in the optimal amount of agents and especially in the optimal contract and skill mix.
Very few utilise more advanced technology, finding the optimal agents pool and determining which agents can best be added to the current pool is hardly done. 

There are no papers solving the pool optimization problem completely. Some papers, such as \cite{AvramidisCGEP,BhulaiPK,LiKooleJouini-omnisim}, as discussed in the previous section, solve the shift scheduling problem for a week or a day, but methods have to be found to extend this to longer periods or to somehow aggregate weekly results to say a year. Furthermore, all forms of shrinkage have to be added.
In our opinion, this is the biggest remaining challenge in WFM, and the only possible solution method we see is a time-consuming simulation-optimization procedure, possibly sped up using ML as in \cite{LiWangKoole-scheduling}.

A simpler solution to the pool composition problem might be to use some rule of thumb. 
Chevalier et al.~\cite{ChevalierST} studies, using approximations based on networks of overflow queues, that 80\% specialized and 20\% fully flexible agents works surprisingly well in many situations. This holds for the staffing problem, random form of shrinkage will likely make the need for flexible agents higher in the pool composition problem.
Also Wallace \& Whitt~\cite{WallaceWhitt-SBR} show, using simulations, that a little flexibility goes a long way, in a situation where agents have 1 or 2 skills and a topology that ``connects" all skills.

\section{Routing and intra-day management}%%%%%%%%%%%%%%%%%%%%%%%%%%%%%%%%%%%%%%%%

Routing in call centers is most of the static: it is entered once in the telephony switch or ACD (Automatic Call Distributor) and it does not depend on current service or staffing levels. Typically, routing is arranged through priorities of agents for certain types of calls, which can be different per agent: agents typically have primary and secondary skills. SLAs can be different for all skills and channels, and priorities which SLAs are most important to be met can be set. When multiple agents with the same priority can handle a call, then usually the one with the longest idle time since the last call is selected. (Note that this rule opens the possibility for the agent to trick the system: by going on a one-second break he or she has again the shortest idle time.) 
When an agent becomes idle, he or she is assigned to the longest waiting call among the highest priority calls. Nowadays, more sophisticated routing rules are supported by ACD like Genesys. For example, threshold policies based on the queue size or customers' waiting time. Customer satisfaction can also be considered by assigning calls to the agent who has the best resolution rate. 
Although this gives many possibilities for routing, and many parameters to be set, there is no guarantee whatsoever that the best possible performance is achieved. For this reason, intra-day managers often change priorities of agents during the day. Unfortunately, they are not supported by software and cannot oversee all implications of their actions which are therefore often highly suboptimal. Systems such as \cite{AvayaBA} try to improve such situations. But experiences are mixed due to a lack of understanding and control by the user.
 Ideally, instead of letting intra-day managers make last-minute adjustments to the system, the routing rules in the ACD should be designed with a full evaluation, validation and optimization. 

Many routing algorithms have been proposed in the literature, but mainly for a heavy-traffic regime with fixed staffing, such as \cite{mandelbaum2004scheduling, milner2008service, atar2005scheduling, atar2010cmu, armony2010fair, ward2013blind}. 
Little studies on SBR exist that take various service levels and also fairness between agents into account. Notable exceptions are Chan et al.~\cite{ChanKE-routing} and Li \& Koole~\cite{LiKoole-routing}, both using simulation. \cite{ChanKE-routing} considers a policy that depends through weights on the service and occupancy levels. The weights that give optimal stationary performance are obtained requiring full knowledge of arrival rates and staffing levels. \cite{LiKoole-routing} is also based on weights, but introduces a heuristic to obtain the best performance by the end of a day without explicitly using the system's parameters but the service level up to that moment, which is the usual performance measure in practice.

Routing between channels is called blending. Most of the studies consider blending inbound and outbound calls. Bhulai \& Koole~\cite{BhulaiK03ieeeac} and Gans \& Zhou~\cite{GansZ03} both show that a non-work-conserving policy is optimal: some agents should be kept free for inbound calls, even though outbound calls are waiting to be handled. Otherwise, the SL on inbound will be too low. This greatly improves the efficiency compared to separate agent groups, and it is robust to changes in parameters such as the arrival rate. Other threshold policies can be found in \cite{pang2015logarithmic, deslauriers2007markov, legros2017reservation}, to name a few.
Legros et al.~\cite{LegrosJouiniKoole-routing} develops a threshold policy that adaptively adjusts the number of agents reserved for inbound calls to achieve the SLA of inbound calls as well as maximise the throughput of emails.

As mentioned in the staffing section, no papers have discussed the blending of asynchronous channels such as inbound and chat. While a few papers deal with the routing of chats (Tezcan and Zhang~\cite{tezcan2014routing}, Legros et al.~\cite{legros2019scheduling}), they both consider single chat type and identical agents. Tezcan and Zhang~\cite{tezcan2014routing} gives a routing rule that minimizes the abandonment rate and the staffing level in the long run. Legros et al.~\cite{legros2019scheduling} considers that customers can also abandon during the service due to long handling times, and propose a routing policy which allows agents to not work up to the maximum number of chats even when the queue is not empty. Further relevant references include Cui and Tezcan~\cite{cui2016approximations} and Luo and Zhang~\cite{luo2013staffing}.

One may notice that the large body of literature mentioned above focuses merely on call center efficiency. Their targets are set to minimize the speed of response, abandonments, SL, and so on. The quality metrics such as call resolution, customer satisfaction and agent preference are barely taken into account. One of the reasons is that the relevant data cannot be easily retrieved from the call center system, requiring extra processing steps. Some exceptions are Ghareeb et al.~\cite{ghareeb2016optimal} and Zhan and Ward~\cite{zhan2014threshold}. 

Intra-day management are changes made to the deployment of agents during the day of execution (or just before). They can be related to the activities they do. Sometimes this is motivated by the SL: agent priorities can be changed, or for example meetings can be cancelled to improve the SL or even scheduled at the last moment when many agents are idle. The changes in activity can also have other motivations, such as the urgent need to schedule a meeting. At all times the consequences to the SL should be taken into account.

Next to changes in activity intra-day management deals with changes in working hours. This starts as soon as the schedule is published by the planners, when for example agents request schedule changes for personal reasons, or when the forecast has changed significantly and more or less agents are needed. This continues throughout the day of execution, many call centers have a flexible workforce layer through which they can up or downscale on a short notice, even during the day itself. The management of this is often not based on SL predictions, and also few papers address this type of issue. An exception is Roubos et al.~\cite{RoubosBK} in which the staffing levels are adapted during the day in an optimal way as to obtain the required SL by the end of the day.

\section{Design}%%%%%%%%%%%%%%%%%%%%%%%%%%%%%%%%%%%%%%%%%%%%%%%%%%%%%%

We start this section with some general guiding principles on how the workforce should be planned.

Decisions that limit flexibility should be taken as late as possible. That way we can better deal with fluctuations, because for all types of fluctuations it holds that over time more information becomes available, i.e., the variability of the unknown variable decreases over time.  E.g., take a multi-skilled call center. During the scheduling phase skills can be assigned to agents blocking them for other skills, unless traffic management changes the schedule. Letting SBR do the assignment is much more efficient, even a fixed assignment at the last moment is better because the latest forecast can be used, and availability is fully known, you known for example who is ill. A re-assignment could be part of intra-day management, but we then schedule in the first place? 

An objection to SBR is that agents have to change skill (e.g., move from one language to another) often, which can be annoying. Similar objections hold against blending, especially when email handling is interrupted for inbound calls. Good routing however can avoid that: in certain systems you can limit the number of times that an email might be interrupted, and one can think of similar solutions for blending.

When the decision is related to something that influences employee satisfaction then making decisions later might be more efficient but at the same time decrease employee satisfaction which negatively impacts the performance of the call center.
However, flexibility is not always required at the maximum level, asking a few fraction of the employees to be flexible might give you the majority of the advantages of flexibility, which is the next guiding principle.

A little flexibility goes a long way. Wallace \& Whitt~\cite{WallaceWhitt-SBR} observed this for the number of multi-skilled agents in an SBR setting, but this holds in general: a few agents with part-time shifts, a few back-office agents who can help in the front-office (the call center), etc. Another way to state it is that flexibility shows decreasing returns. From this it also follows that you can better have a bit of multiple types of flexibility, than a large amount of one type of flexibility. 
However, using all these forms of flexibility together in the smartest way is a challenging task that requires tooling, nobody can immediately grasp the consequences of one agent less on all skills, even the one he or she does not have. That brings us to the final guideline.  

Automated decision making is preferred over manual. Few decisions are fully manual or automated, for most decisions there is some tool (which can be a spreadsheet) that supports the decision. The better the tooling, the less human interference is required. We argue that a higher agree of automation is usually better: advanced knowledge in the form of algorithms can be implemented in software, knowledge that most planners will never obtain. An important constraint is the outcomes should be transparent, to be able to explain the outcomes and interact with it. As an example, take an agent who wants the afternoon off. Usually an intra-day manager looks at the current SL, and makes a decision on the basis of that. It would be much better to have a SL prediction for the rest of the day to base the decision on, but that requires advanced tooling. From there it is a small step to an automated system that compares the predicted SL with the SLA and that makes a decision on that basis. Clearly this makes the call center more efficient but also makes the WFM smaller, leading to additional costs savings which are usually higher than the software costs. 

When designing a call center WFM is only one of the aspects that have to be taken into account, each decision should also be evaluated from the point of view of WFM: what are the consequences for the SL, the agent satisfaction, and the efficiency, i.e., the costs? Quite often WFM is not in the loop when design decisions are made. E.g., when the decision is made not to look at AHT anymore to allow agents to give the best possible service, what are the consequences on the required workforce if the AHT gets much longer? It might be customer-friendly to expand the opening hours, but what are the consequences for the SL? There are many questions of this type. The tools described in the previous sections can often be used to solve these problems.

There are different examples in the literature of designs that are well-built, also from the point of view of WFM. We cite a few of them.
Jouini et al.~\cite{JouiniDR-teams} studied a multi-skilled situation with different teams where every team has its own customer base. To obtain flexibility and with that economies of scale unidentified callers are routed to the least occupied team. This combines motivational incentives, such as being able to compare teams on quality-of-service and up-sell, with WFM aspects. 
Legros et al.~\cite{legros2015flexible} proposed a new SBR architecture for the situation that every agent has two skills per agent. The efficiency of the proposed architecture is compared to chaining using simulation.
Ak{\c s}in et al.~\cite{akcsin2008call} discussed the optimal capacity levels under two different outsourcing contract models: volume-based and capacity-based contracts. They observed that no contract type is universally preferred, and both the operating environments and cost-revenue structures matter. Gans and Zhou~\cite{gans2007call}, Hasija et al.~\cite{hasija2008call}, and Gurvich and Perry~\cite{gurvich2012overflow} studied the overflow operating rules in the outsourcing environment. Some call centers offer a call-back option to smooth the arrival traffic, whereby customers may register a request when all agents are busy. Later the system will call them back within a pre-specified time slot. In this way, waiting inbound turns to the outbound task at scheduled moments. The analysis of such systems is conducted in \cite{armony2004contact, armony2004customer, hathaway2020don, legros2016optimal, legros2017call}.

\paragraph{Acknowledgements} We are grateful to the stimulating environment that CCmath offers and that made this paper and its research possible. We are especially grateful to Giuseppe Catanese, Alex Roubos and Wout Bakker for their feedback. CCmath has its own algorithms for forecasting, staffing and scheduling of which we were not allowed to disclose the details for commercial reasons. Note that this overview might be biased towards the situation found at CCmath's clients. 

The second author also wishes to thank the Vrije Universiteit for the hospitality that was offered to her over multiple years.

\bibliographystyle{plain}
\bibliography{refs} 

\end{document}